\newif\iftr\trtrue

\documentclass{sig-alternate-2013}
\usepackage{array}
\usepackage{fixltx2e}
\usepackage{textcomp}

\usepackage{amsthm}
\usepackage{bm}
\usepackage{pl}
\usepackage{stmaryrd}
\usepackage{xcolor}
\usepackage{tikz}
\usetikzlibrary{arrows,positioning,shapes,shapes.geometric,patterns}
\usepackage{url}
\usepackage{float}
\usepackage{cite}
\usepackage{hyperref}
\hypersetup{
   pdftitle={Belief Semantics of Authorization Logic},
   pdfauthor={Andrew K. Hirsch and Michael R. Clarkson}
 }
\usepackage{defs}
\usepackage{naldefs}

\iftr
\permission{}
\copyrightetc{Copyright 2013 Andrew K. Hirsch and Michael R. Clarkson}
\else
\newfont{\mycrnotice}{ptmr8t at 7pt}
\newfont{\myconfname}{ptmri8t at 7pt}
\let\crnotice\mycrnotice%
\let\confname\myconfname%
\permission{Permission to make digital or hard copies of all or part of this work for personal or classroom use is granted without fee provided that copies are not made or distributed for profit or commercial advantage and that copies bear this notice and the full citation on the first page. Copyrights for components of this work owned by others than ACM must be honored. Abstracting with credit is permitted. To copy otherwise, or republish, to post on servers or to redistribute to lists, requires prior specific permission and/or a fee. Request permissions from permissions@acm.org.}
\conferenceinfo{CCS'13,}{November 4--8, 2013, Berlin, Germany.}
\copyrightetc{Copyright 2013 ACM \the\acmcopyr}
\crdata{978-1-4503-2477-9/13/11\ ...\$15.00.\\
http://--enter the whole DOI string from rightsreview form confirmation}
\fi

\begin{document}

\pagestyle{plain}

\title{Belief Semantics of Authorization Logic}

\numberofauthors{2}
\author{
Andrew K. Hirsch \\
       \affaddr{Department of Computer Science}\\
       \affaddr{George Washington University}\\
       \affaddr{Washington, D.C., United States}\\
       \email{akhirsch@gwu.edu}
\alignauthor
Michael R. Clarkson \\
       \affaddr{Department of Computer Science}\\
       \affaddr{George Washington University}\\
       \affaddr{Washington, D.C., United States}\\      
       \email{clarkson@gwu.edu}
}

\maketitle

\begin{abstract}
Authorization logics have been used in the theory of computer security to reason about access control decisions.
In this work, a formal belief semantics for authorization logics is given.
The belief semantics is proved to subsume a standard Kripke semantics.
The belief semantics yields a direct representation of principals' beliefs, without resorting to the technical machinery used in Kripke semantics.
A proof system is given for the logic; that system is proved sound with respect to the belief and Kripke semantics.  
The soundness proof for the belief semantics, and for a variant of the Kripke semantics, is mechanized in Coq.
\end{abstract}



\section{Introduction}
\label{sec:intro}

Authorization logics are used in computer security to reason about whether \emph{principals}---computer or human agents---are permitted to take actions in computer systems.  
The distinguishing feature of authorization logics is their use of a $\NALSAYS$ connective:
intuitively, if principal $p$ believes that formula $\phi$ holds, then formula $\nalSays{p}{\phi}$ holds.  
Access control decisions can then be made by reasoning about (i) the beliefs of principals, (ii) how those beliefs can be combined to derive logical consequences, and (iii) whether those consequences entail \emph{guard formulas}, which must hold for actions to be permitted.

Many systems that employ authorization logics have been proposed~\cite{LampsonABW92,WobberABL94,AppelF99,LiGF00,Jim01,DeTreville02,LiMW02,BeckerS04,BauerGMRRR05,FournetGM05,PimlottK06,CederquistCDEHL07,CirilloJPR07,Lesniewski-LaasFSMK07,GurevichN08,JiaVMZZSZ08,BeckerFG10,SirerBRSWWS11}, but few authorization logics have been given a formal semantics~\cite{AbadiBLP93,Howell00,GargA08,Garg08,GenoveseGR12}.  
Though semantics might not be immediately necessary to deploy authorization logics in real systems, 
semantics yield insight into the meaning of formulas, and 
semantics enable proof systems to be proved sound---which might require proof rules and axioms to be corrected, if there are any lurking errors in the proof system.

For the sake of security, it is worthwhile to carry out such soundness proofs.  
Given only a proof system, we must trust that the proof system is correct.  
But given a proof system and a soundness proof, which shows that any provable formula is semantically valid, we now have evidence that the proof system is correct, hence trustworthy. 
The soundness proof thus relocates trust from the proof system to the  proof itself---as well as to the semantics, which ideally offers more intuition about formulas than the proof system itself.

Semantics of authorization logics are usually based on \emph{possible worlds}, as used by Kripke~\cite{Kripke63}.
\emph{Kripke semantics} posit an indexed \emph{accessibility relation} on possible worlds.
If at world $w$, principal $p$ considers world $w'$ to be possible, then $(w,w')$ is in $p$'s accessibility relation.
We denote this as $w \leq_p w'$.
Authorization logics use Kripke semantics to give meaning to the $\NALSAYS$ connective:
semantically, $\nalSays{p}{\phi}$ holds in a world $w$ iff for all worlds $w'$ such that $w \leq_p w'$, formula $\phi$ holds in world $w'$.
Hence a principal says $\phi$ iff $\phi$ holds in all worlds the principal considers possible.%
\footnote{The $\NALSAYS$ connective is, therefore, closely related to the modal necessity operator $\Box$~\cite{HughesC96} and the epistemic knowledge operator $K$~\cite{FaginHMV95}.}

The use of Kripke semantics in authorization logic thus requires installation of possible worlds and accessibility relations into the semantics, solely to give meaning to $\NALSAYS$.  
That's useful for studying properties of logics and for building decision procedures.
But, unfortunately, it doesn't seem to correspond to how principals reason in real-world systems.
Rather than explicitly considering possible worlds and relations between them, principals typically begin with some set of base formulas they believe to hold---perhaps because they have received digitally signed messages encoding those formulas, or perhaps because they invoke system calls that return information---then proceed to reason from those formulas.
So could we instead stipulate that each principal $p$ have a set of beliefs $\worldview(p)$, called the \emph{worldview} of $p$, such that $\nalSays{p}{\phi}$ holds iff $\phi \in \worldview(p)$?
That is, a principal says $\phi$ iff $\phi$ is in the worldview%
\footnote{Worldviews were first employed by NAL~\cite{SchneiderWS11}, which pioneered an informal semantics based on them.}%
of the principal?

This paper answers that question in the affirmative.  
We give two semantics for an authorization logic: a Kripke semantics (\S\ref{sec:kripkesemantics}), and a new \emph{belief semantics} (\S\ref{sec:beliefsemantics}), which employs worldviews to interpret $\NALSAYS$.%
\footnote{Our belief models are an instance of the \emph{syntactic} approach to modeling knowledge~\cite{Eberle74,MooreH79,FaginHMV95,Konolige86}.}  
We show (\S\ref{sec:semanticequiv}) that belief semantics subsume Kripke semantics, in the sense that a belief model can be constructed from any Kripke model. 
A formula is valid in the Kripke model iff it is valid in the constructed belief model.
As a result, the technical machinery of Kripke semantics can be replaced by belief semantics.
This potentially increases the trustworthiness of an authorization system, because the semantics is closer to how principals reason in real systems.

The particular logical system we introduce in this paper is FOCAL, First-Order
Constructive Authorization Logic.  
FOCAL extends a well-known authorization logic, cut-down dependency core calculus (CDD)~\cite{Abadi07}, from a propositional language to a language with first-order functions and relations on system state.  
Functions and relations are essential for reasoning about authorization in a real operating system---as exemplified in Nexus Authorization Logic (NAL)~\cite{SchneiderWS11}, of which FOCAL and CDD are both fragments.

Having given two semantics for FOCAL, we then turn to the problem of proving soundness.  
It turns out that the NAL proof system is unsound with respect to\ the semantics presented here:  
NAL allows derivation of a well-known formula (cf.~\S\ref{sec:focalevsnal}) that  our semantics deems invalid.  
A priori, the fault could lie with our semantics or with NAL's proof system.
However, if the logic is to be used in a distributed setting without globally-agreed upon state, then the proof system should not allow the formula to be derived.  
So if NAL is to be used in such settings, its proof system needs to be corrected.
CDD is also unsound with respect to our semantics. 
However, CDD has been proved sound with respect to a different semantics~\cite{GargA08}.
This seeming discrepancy---sound vs.\ unsound---illuminates a previously unexplored difference (cf.\ \S\ref{sec:focalevscdd}) between how NAL and CDD interpret $\NALSAYS$.

To achieve soundness for FOCAL, we develop a revised proof system; the key technical change is using localized hypotheses in the proof rules.
In \S\ref{sec:proofsystem}, we prove the soundness of our proof system with respect to both our belief and Kripke semantics.  
This result yields the first soundness proof with respect to belief semantics for an authorization logic.

Having relocated trust into the soundness proof, we then seek a means to increase the trustworthiness of that proof.  
We formalize the syntax, proof system, belief semantics, and  Kripke semantics in the Coq proof assistant,\footnote{\url{http://coq.inria.fr}} and we mechanize the proofs of soundness for both the belief semantics and the Kripke semantics.  
That mechanization relocates trust from our soundness proof to Coq, which is well-studied and is the basis of many other formalizations.
Our Coq formalization contains about 2,400 lines of code.%
\footnote{Our implementation is available from \url{http://faculty.cs.gwu.edu/~clarkson/projects/focal/}.}

This paper thus advances the theory of computer security 
with the following novel contributions:
\begin{itemize}
\item the first formal belief semantics for authorization logic,
\item a proof of equivalence between belief semantics and Kripke semantics,
\item a proof system that is sound with respect to\ belief and Kripke semantics, and
\item the first machine-checked proof of soundness for an authorization-logic proof system.
\end{itemize}

We proceed as follows.  
\S\ref{sec:beliefsemantics} presents FOCAL and its belief semantics.  
\S\ref{sec:kripkesemantics} gives a Kripke semantics for FOCAL.
\S\ref{sec:semanticequiv} proves the relationship of the belief semantics to the Kripke semantics.
\S\ref{sec:proofsystem} gives a proof system for FOCAL and proves its soundness with respect to\ the Kripke semantics.
\S\ref{sec:relatedwork} discusses related work, and \S\ref{sec:conclusion} concludes.
All proofs appear in the appendix.

\section{Belief Semantics}
\label{sec:beliefsemantics}

FOCAL is a constructive, first-order, multimodal logic. 
The key features that distinguish it as an authorization logic are the $\NALSAYS$ and $\NALSPEAKSFOR$ connectives, invented by Lampson et al.~\cite{LampsonABW92}.
These are used to reason about authorization---for example, access control in a distributed system can be modeled in the following standard way:
\begin{example}\hspace*{0pt}
A \emph{guard} implements access control for a print\-er $p$.  
To permit printing to $p$, the guard must be convinced that guard formula $\nalSays{\mathit{PrintServer}}{\mathit{printTo}(p)}$ holds, where $\mathit{PrintServer}$ is the principal representing the server process.  
That formula means $\mathit{PrintServer}$ believes $\mathit{printTo}(p)$ holds.
To grant printer access to user $u$, the print server can issue the statement $\nalSpeaksfor{u}{\mathit{PrintServer}}$.  
That formula means anything $u$ says, the $\mathit{PrintServer}$ must also say.
So if $\nalSays{u}{\mathit{printTo}(p)}$, then $\nalSays{\mathit{PrintServer}}{\mathit{printTo}(p)}$, which satisfies the guard formula hence affords the user access to the printer.
\end{example}

\begin{figure}
\begin{align*} 
\tau ::= 
\quad&x \bnf f(\tau, \ldots, \tau) \\
\phi ::= 
\quad&\nalTrue \bnf \nalFalse \bnf r(\tau, \ldots, \tau) \bnf \tau_1 = \tau_2 \\
\bnf &\phi_1 \nalAnd \phi_2 \bnf \phi_1 \nalOr \phi_2 \bnf \phi_1 \nalImplies \phi_2 \bnf \nalNot \phi \\
\bnf &\nalForall{x}{\phi} \bnf \nalExists{x}{\phi} \\
\bnf &\nalSays{\tau}{\phi} \bnf \nalSpeaksfor{\tau_1}{\tau_2}
\end{align*}
\caption{Syntax of FOCAL\label{fig:syntax}}
\end{figure}

Figure~\ref{fig:syntax} gives the formal syntax of FOCAL. 
There are two syntactic classes, terms $\tau$ and formulas $\phi$.  
Metavariable $x$ ranges over first-order variables, $f$ over first-order functions, and $r$ over first-order relations.  

Formulas of FOCAL do not permit monadic second-order universal quantification, unlike CDD and NAL.
In NAL, that quantifier was used only to define $\nalFalse$ and $\NALSPEAKSFOR$ as syntactic sugar.
FOCAL instead adds these as primitive connectives to the logic.  
FOCAL also defines $\nalNot \phi$ as a primitive connective, but it could equivalently be defined as syntactic sugar for $\phi \nalImplies \nalFalse$.

Syntactically, FOCAL is thus CDD without second-order quantification, but with first-order terms and quantification and a primitive $\NALSPEAKSFOR$ connective. 
Likewise, FOCAL is NAL without second-order quantification, subprincipals, group principals, and restricted delegation, but with a primitive $\NALSPEAKSFOR$ connective. 

\subsection{Semantic models}
\label{sec:beliefmodels}

The belief semantics of FOCAL combines first-order constructive models with worldviews, which are used to interpret $\NALSAYS$ and $\NALSPEAKSFOR$.
To our knowledge, this semantics is new in the study of authorization logics.
Our presentation mostly follows the semantics of intuitionistic predicate calculus given by Troelstra and van Dalen~\cite{TroelstravD88a}.

\paragraph*{First-order models} 
A \emph{first-order model with equality} is a tuple $(D, \mathord{=}, R, F)$.
The purpose of a first-order model is to interpret the first-order fragment of the logic, specifically first-order quantification, functions, and relations.  
$D$ is a set, the \emph{domain} of individuals.  
Semantically, quantification in the logic ranges over these individuals.
$R$ is a set $\setdef{r_i}{i \in I}$ of relations on $D$, indexed by set $I$.
Likewise, $F$ is a set $\setdef{f_j}{j \in J}$ of functions on $D$, indexed by set $J$.
There is a distinguished equality relation $=$, which is an equivalence relation on $D$, such that equal individuals are indistinguishable by relations and functions.

To interpret first-order variables, the semantics employs $\emph{valuation}$ functions, which map variables to individuals.  
We write $v(x)$ to denote the individual that variable $x$ represents in valuation $v$.
And we write $v\subst{d}{x}$ to denote the valuation that is the same as $v$ except that $v(x)=d$.

\paragraph*{Constructive models}
A \emph{constructive model} is a tuple $(W,$ $\mathord{\leq}, s)$.  
The purpose of constructive models is to extend first-order models to  interpret the constructive fragment of the logic, specifically implication and universal quantification.
$W$ is a set, the possible worlds. 
We denote an individual world as $w$.  
Intuitively, a world $w$ represents the state of knowledge of a constructive reasoner.
\emph{Constructive accessibility relation} $\leq$ is a partial order on $W$. 
If $w \leq w'$, then the constructive reasoner's state of knowledge could grow from $w$ to
$w'$. 
But unlike in classical logic, the reasoner need not commit to a formula $\phi$ being either true or false at a world. 
Suppose that at world $w'$, where $w \leq w'$, the reasoner concludes that $\phi$ holds.  
And at world $w''$, where $w \leq w''$, the reasoner concludes that $\nalNot\phi$ holds.
But at world $w$, the reasoner has not yet concluded that either $\phi$ or $\nalNot\phi$ holds.
Then Excluded Middle ($\phi \nalOr \nalNot \phi$) doesn't hold at $w$. 

Function $s$ is the \emph{first-order interpretation function}. 
It assigns a first-order model $({D_w}, {=_w}, {R_w}, {F_w})$ to each world $w$.
Let the individual elements of $R_w$ be denoted as $r_{i,w}$, and the elements of $F_w$ as $f_{j,w}$.
Thus, $s$ enables a potentially different first-order interpretation at each world.  
But to help ensure that the constructive reasoner's state of knowledge only grows---hence never invalidates a previously admitted construction---we require $s$ to be monotonic w.r.t.\ $\leq$.  
That is, if $w \leq w'$ then (i) $D_w \subseteq D_{w'}$, (ii) $d =_w d'$ implies $d =_{w'} d'$, (iii) $r_{i,w} \subseteq r_{i,w'}$, and (iv) for all tuples $\vec{d}$ of individuals in $D_w$, it holds that $f_{j,w}(\vec{d}) =_w f_{j,w'}(\vec{d})$.

It's natural to wonder why we chose to introduce possible worlds into the semantics here after arguing against them in \S\ref{sec:intro}.
Note, though, that the worlds in the constructive model are being used to model only the constructive reasoner---which we might think of as the guard, who exists outside the logic and attempts to ascertain the truth of formulas---not any of the principals reasoned about inside the logic.
Moreover, we have not introduced any accessibility relations for principals, but only a single accessibility relation for the constructive reasoner.  
So the arguments in \S\ref{sec:intro} don't apply.
It would be possible to eliminate our usage of possible worlds by employing a \emph{Heyting algebra semantics}~\cite{TroelstravD88b} of constructive logic.  
But possible worlds blend better with the Kripke semantics in \S\ref{sec:kripkesemantics}.

It's also natural to wonder why FOCAL is constructive rather than classical.  
Schneider et al.~\cite{SchneiderWS11} write that constructivism preserves evidence:
``Constructive logics are well suited for reasoning about authorization{\ldots}because constructive proofs include all of the evidence used for reaching a conclusion and, therefore, information about accountability is not lost. 
Classical logics allow proofs that omit evidence.''  
They argue that Excluded Middle, used as an axiom in a proof, would omit evidence by failing to indicate whether access was granted on the basis of $\phi$ holding or $\nalNot\phi$ holding. 
Garg and Pfenning~\cite{GargP06} also champion the notion of evidence in authorization logics, writing that ``[constructive logics] keep evidence contained in proofs as direct as possible.''
Regardless, we believe that a classical version of FOCAL could be created without difficulty.

\paragraph*{Belief models}
A \emph{belief model} is a tuple $(W, \leq, s, P, \omega)$.  
The purpose of belief models is to extend constructive models to  interpret $\NALSAYS$ and $\NALSPEAKSFOR$. 
The first part of a belief model, $(W, \leq, s)$, must itself be a constructive model.  
The next part, $P$, is the set of principals.  
Although individuals can vary from world to world in a model, the set of principals is fixed across the entire model.
Assuming a fixed set of principals is consistent with other authorization logics~\cite{GargA08,GenoveseGR12,Garg08}, with constructive multimodal logics~\cite{Wijesekera90,Simpson94} (which have a fixed set of modalities), and with classical multimodal epistemic logics~\cite{FaginHMV95} (which have an indexed set modalities, typically denoted $K_i$, where the index set is fixed)---even though constructivist philosophy might deem it more sensible to allow $P$ to grow with $\leq$.

Because we make no syntactic distinction between individuals and principals, all principals must also be individuals:  $P$ must be a subset of $D_w$ for every $w$.
First-order quantification can therefore range over individuals as well as principals.
For example, to quantify over all principals, we can write $\nalForall{x}{\mathit{IsPrin}(x) \nalImplies \phi}$, where $\mathit{IsPrin}$ is a relation that holds for all $x \in P$.
Nonetheless, this does not constitute truly intuitionistic quantification, because the domain of principals is constant.  
Quantification over a non-constant domain of principals is theoretically of interest, but we know of no authorization logic that has used it.  

We define an equality relation $\peq$ on principals, such that principals are equal iff they are equal at all worlds.  Formally, $p \peq p'$ iff, for all $w$, it holds that $p =_w p'$.

The final part of a belief model, worldview function $\omega$, yields the beliefs of a principal $p$:
the set of formulas that $p$ believes to hold in world $w$ under first-order valuation $v$ is $\worldview(w,p,v)$.
For sake of simplicity, \S\ref{sec:intro} used notation $\worldview(p)$ when first presenting the idea of worldviews.  
Now that we're being precise, we also include $w$ and $v$ as arguments.
To ensure that the constructive reasoner's knowledge grows monotonically, worldviews must be monotonic w.r.t.\ $\leq$:
\begin{mrcdefn}{Worldview Monotonicity}
If $w \leq w'$ then $\worldview(w, p, v)$ $\subseteq \worldview(w', p, v)$.
\end{mrcdefn}
To ensure that whenever principals are equal they have the same worldview, we require the following:
\begin{mrcdefn}{Worldview Equality} 
If $p \peq p'$, then, for all $w$ and $v$, it holds that $\worldview(w, p, v) = \worldview(w, p', v)$.
\end{mrcdefn}
And we also require the following conditions to ensure that valuations cannot cause worldviews to distinguish alpha-equivalent formulas:
\begin{mrcdefn}{Worldview Valuations}
\vspace{-4pt}
\begin{enumerate}
\item If $x \notin \FV(\phi)$ 
   then $\phi \in \worldview(w,p,v)$  iff, for all $d \in D_w$, it holds that $\phi \in \worldview(w,p,v\subst{d}{x})$.
\item If $x \in \FV(\phi)$ and $y \notin \FV(\phi)$
   then, for all $d \in D_w$, it holds that $\phi \in \worldview(w,p,v\subst{d}{x})$  iff  $\phi\subst{y}{x} \in \worldview(w,p,v\subst{d}{y})$, where $\phi\subst{y}{x}$ denotes the capture-avoiding substitution of $y$ for $x$ in formula $\phi$.
\end{enumerate}
\end{mrcdefn}
Condition (1) ensures that if $x$ is irrelevant to $\phi$, then the value of $x$ is also irrelevant to whether $p$ believes $\phi$.  
Condition (2) ensures that if $x$ is relevant to $\phi$, then only its value---not its name---is relevant to whether $p$ believes $\phi$.

\subsection{Semantic validity}
\label{sec:beliefsemanticvalidity}

\begin{figure*}
\begin{equation*}
\begin{array}{lcl}
B,w,v \models \nalTrue & & \text{always}  \\
B,w,v \models \nalFalse & & \text{never}  \\
B,w,v \models r_i(\vec{\tau}) & \text{iff} & \mu(\vec{\tau}) 
  \in r_{i,w} \\
B,w,v \models \tau_1 = \tau_2 & \text{iff} & \mu(\tau_1) =_w
  \mu(\tau_2) \\
B,w,v \models \phi_1 \nalAnd \phi_2 & \text{iff} & 
  B,w,v \models \phi_1 \text{~and~}
  B,w,v \models \phi_2 \\
B,w,v \models \phi_1 \nalOr \phi_2 & \text{iff} & 
  B,w,v \models \phi_1 \text{~or~}
  B,w,v \models \phi_2 \\
B,w,v \models \phi_1 \nalImplies \phi_2 & \text{iff} & 
  \text{for all~} w' \geq w: B,w',v \models \phi_1 \text{~implies~}  B,w',v \models \phi_2 \\
B,w,v \models \nalNot \phi & \text{iff} & 
  \text{for all~} w' \geq w: B,w',v \not\models \phi \\
B,w,v \models \nalForall{x}{\phi} & \text{iff} & 
  \text{for all~} w' \geq w, d \in D_{w'}: 
  B,w',v\subst{d}{x} \models \phi \\
B,w,v \models \nalExists{x}{\phi} & \text{iff} & 
  \text{there exists~} d \in D_{w}: 
  B,w,v\subst{d}{x} \models \phi \\
B,w,v \models \nalSays{\tau}{\phi} & \text{iff} & 
  \phi \in \worldview(w, \mu(\tau), v) \\ 
B,w,v \models \nalSpeaksfor{\tau_1}{\tau_2} & \text{iff} & 
  \text{for all~} w' \geq w: \worldview(w', \mu(\tau_1), v) \subseteq \worldview(w', \mu(\tau_2), v)
\end{array}
\end{equation*}
\caption{FOCAL validity judgment for belief semantics\label{fig:focalbeliefsemantics}}
\end{figure*}

Figure~\ref{fig:focalbeliefsemantics} gives a belief semantics of FOCAL.
The validity judgment is written $B, w, v \models \phi$ where $B$ is a belief model and $w$ is a world in that model.
As is standard, $B \models \phi$ holds iff, for all $w$ and $v$, it holds that $B, w, v \models \phi$; whenever $B \models \phi$, then $\phi$ is a \emph{necessary} formula in model $B$.
And $B,v \models \phi$ holds iff for all $w$, it holds that $B,w,v \models \phi$; whenever $B,v \models \phi$, then $\phi$ is a \emph{valuation-necessary} formula.
Likewise, $\models \phi$ holds iff, for all $B$, it holds that $B \models \phi$; and whenever $\models \phi$, then $\phi$ is a \emph{validity}.
Let $B,w,v \models \Gamma$, where $\Gamma$ is a set of formulas, denote that for all $\psi \in \Gamma$, it holds that $B,w,v \models \psi$.
Finally, $\Gamma \models \phi$ holds iff, for all $B$, $w$, and $v$, it holds that $B,w,v \models \Gamma$ implies $B,w,v \models \phi$; whenever $\Gamma \models \phi$, then $\phi$ is a \emph{logical consequence} of $\Gamma$.

The semantics relies on an auxiliary \emph{interpretation} function $\mu$ that maps syntactic terms $\tau$ to semantic individuals:
\begin{align*}
\mu(x) & = v(x) \\
\mu(f_j(\vec{\tau})) & = f_{j,w}(\mu(\vec{\tau})) 
\end{align*}
Implicitly, $\mu$ is parameterized on belief model $B$, world $w$, and valuation $v$, but for notational simplicity we omit writing these as arguments to $\mu$ unless necessary for disambiguation.
Variables $x$ are interpreted by looking up their value in $v$;
functions $f_j$ are interpreted by applying their first-order interpretation $f_{j,w}$ at world $w$ to the interpretation of their arguments.  
Notation $\vec{\tau}$ represents a list $\tau_1, \tau_2, \ldots, \tau_n$ of terms.
And $\mu(\vec{\tau})$ denotes the pointwise application of $\mu$ to each element of that list, producing $\mu(\tau_1), \ldots, \mu(\tau_n)$.

The first-order, constructive fragment of the semantics is routine.
The semantics of $\NALSAYS$ is the intuitive semantics we wished for in \S\ref{sec:intro}:
A principal $\mu(\tau)$ says $\phi$ exactly when $\phi$ is in that principal's worldview $\worldview(w, \mu(\tau), v)$.  
And a principal $\mu(\tau_1)$ speaks for another principal $\mu(\tau_2)$ exactly when, in all constructively accessible worlds, everything $\mu(\tau_1)$ says, $\mu(\tau_2)$ also says.

Note that some syntactic terms may represent individuals that are not principals.
For example, the integer $42$ is presumably not a principal in $P$, but it could be an individual in some domain $D_w$.  
An alternative would be to make FOCAL a two-sorted logic, with one sort for individuals and another sort for principals.
Instead, we allow individuals who aren't principals to have beliefs, because it simplifies the definition of the logic.  
The worldviews of non-principal individuals contain all formulas.  
Formally, for any individual $d$ such that $d\not\in P$, and for any world $w$, valuation $v$, and formula $\phi$, it holds that $\phi \in \worldview(w, d, v)$.

We impose a few \emph{well-formedness} conditions on worldviews in this semantics, 
in addition to Worldview Monotonicity and Worldview Equality.
Worldviews must be \emph{closed under logical consequence}---that is, principals must believe all the formulas that are a consequence of their beliefs. 
\begin{mrcdefn}{Worldview Closure}
If $\Gamma \subseteq \worldview(w, p, v)$ and $\Gamma \models \phi$, then $\phi \in \worldview(w, p, v)$.
\end{mrcdefn}
Worldview Closure means that principals are \emph{fully logically omniscient}~\cite{FaginHMV95}.  With its known benefits and flaws~\cite{Parikh87,Stalnaker91}, this has been a standard assumption in authorization logics since their inception~\cite{LampsonABW92}.

The remaining well-formedness conditions are optional, in the sense that they are necessary only to achieve soundness of particular proof rules in \S\ref{sec:proofsystem}.  
Eliminate those rules, and the following conditions would be eliminated.

Worldviews must ensure that $\NALSAYS$ is a \emph{transparent} modality. That is, for any principal $p$, it holds that $\nalSays{p}{\phi}$ exactly when $\nalSays{p}{(\nalSays{p}{\phi})}$:
\begin{mrcdefn}{Says Transparency}
$\phi \in \worldview(w, \mu(\tau), v)$ iff $\nalSays{\tau}{\phi} \in \worldview(w, \mu(\tau), v)$.
\end{mrcdefn}
So $\NALSAYS$ supports \emph{positive introspection}:  
if $p$ believes that $\phi$ holds, then $p$ is aware of that belief, therefore $p$ believes that $p$ believes that $\phi$ holds.  
The converse of that holds as well. 
Recent authorization logics include transparency~\cite{Abadi08,SchneiderWS11}, and it is well known (though sometimes vigorously debated) in epistemic logic~\cite{Hintikka62,HughesC96}.
Says Transparency corresponds to rules \ruleName{says-li} and \ruleName{says-ri} in figure~\ref{fig:focalproofsystem}.

Worldviews must enable principals to delegate, or \emph{hand-off}, to other principals:
if a principal $q$ believes that \linebreak $\nalSpeaksfor{p}{q}$, it should hold that $p$ does speak for $q$.
Hand-off, as the following axiom, existed in the earliest authorization logic~\cite{LampsonABW92}:
\begin{equation}\label{eq:handoff}
(\nalSays{q}{(\nalSpeaksfor{p}{q})}) \nalImplies (\nalSpeaksfor{p}{q})
\end{equation}

\noindent To support it, we adopt a condition that ensures whenever $q$ believes $p$ speaks for $q$, then it really does:
\begin{mrcdefn}{Belief Hand-off}
If $(\nalSpeaksfor{p}{q}) \in \worldview(w, q, v)$ then $\worldview(w, p, v) \subseteq \worldview(w, q, v)$.\end{mrcdefn}
Belief Hand-off corresponds to rule \ruleName{sf-i} in figure~\ref{fig:focalproofsystem}.

\section{Kripke Semantics}
\label{sec:kripkesemantics}

\newif\ifframecond\framecondtrue

\newcommand{\RA}[2]{\ensuremath{\mathord{\leq^{#2}_{\mu(#1)}}}}
\begin{figure*}
\begin{equation*}
\renewcommand{\arraystretch}{1.1}
\begin{array}{lcl}
K,w,v \models \nalSays{\tau}{\phi} & \text{iff} & 
  \text{for all~} w', w'' :
    w \leq w' \leq_{\mu(w',\tau)} w'' \text{~implies~}  K,w'',v \models \phi \\
K,w,v \models \nalSpeaksfor{\tau_1}{\tau_2} & \text{iff} & 
  \RA{\tau_1}{w} \supseteq \RA{\tau_2}{w} \\
K,w,v \models \ldots & \text{iff} & \textit{same as figure~\ref{fig:focalbeliefsemantics}, but substituting K for B}
\end{array}
\end{equation*}
\caption{FOCAL validity judgment for Kripke semantics\label{fig:focalkripkesemantics}}
\end{figure*}

The Kripke semantics of FOCAL combines first-order constructive models with modal (Kripke) models\cite{FaginHMV95,Simpson94,HughesC96}.  
Similar semantic models have been explored before (see, e.g., \cite{Wijesekera90,GenoveseGR12,Garg08}).
Indeed, the only non-standard part of our semantics is the treatment of $\NALSPEAKSFOR$, and that part turns out to be a generalization of previous classical semantics.
Nonetheless, we are not aware of any authorization logic semantics that is equivalent to or subsumes our semantics. 
First-order and constructive models were already presented in \S\ref{sec:beliefsemantics}, so we begin here with modal models.

\subsection{Modal models} 
\label{sec:modalmodels}
A \emph{modal model} is a tuple $(W, \leq, s, P, A)$.
The purpose of modal models is to extend constructive models to interpret $\NALSAYS$ and $\NALSPEAKSFOR$.  
The first part of a modal model, $(W, \leq, s)$, must itself be a constructive model.
The next part, $P$, is the set of principals. 
As with belief models, all principals must be individuals, so $P$ must be a subset of $D_w$ for every $w$. 
Principal equality relation $\peq$ is defined just as in belief models.
The final part of a modal model, $A$, is a set $\setdef{\mathord{\leq_p}}{p \in P}$ of binary relations on $W$, called the \emph{principal accessibility relations}.%
\footnote{In our notation, an unsubscripted $\leq$ always denotes the constructive relation, and a subscripted $\leq$ always denotes a principal relation.}
If $w \leq_p w'$, then at world $w$, principal $p$ considers world $w'$ possible. 
To ensure that equal principals have the same beliefs, we require 
\begin{mrcdefn}{Accessibility Equality}
If $p \peq p'$, then $\mathord{\leq_p} = \mathord{\leq_{p'}}$.
\end{mrcdefn}
Like $\leq$ in a constructive model, we require $s$ to be monotonic w.r.t.\ each $\leq_p$.
This requirement enforces a kind of constructivity on each principal $p$, such that from a world in which individual $d$ is constructed, $p$ cannot consider possible any world in which $d$ has not been constructed.
Unlike $\leq$, none of the $\leq_p$ are required to be partial orders:  
they are not required to satisfy reflexivity, anti-symmetry, or transitivity.

That non-requirement raises an important question.  
In epistemic logics, the properties of what we call the ``principal accessibility relations'' determine what kind of knowledge is modeled~\cite{FaginHMV95}. 
If, for example, these relations must be reflexive, then the logic models \emph{veridical} knowledge: 
if $\nalSays{p}{\phi}$, then $\phi$ indeed holds.  
But that is not the kind of knowledge we seek to model with FOCAL, because principals may say things that in fact do not hold.
So what are the right properties, or \emph{frame conditions}, to require of our principal accessibility relations?
We briefly delay presenting them, so that we can present the Kripke semantics.

\subsection{Semantic validity} 

Figure~\ref{fig:focalkripkesemantics} gives a Kripke semantics of FOCAL.
The validity judgment is written $K, w, v \models \phi$ where $K$ is a modal model and $w$ is a world in that model.
Only the judgments for the $\NALSAYS$ and $\NALSPEAKSFOR$ connectives are given in figure~\ref{fig:focalkripkesemantics}.
For the remaining connectives, the Kripke semantics is the same as the belief semantics in figure~\ref{fig:focalbeliefsemantics}.
Interpretation function $\mu$ remains unchanged from \S\ref{sec:beliefsemantics}, except that it is now implicitly parameterized on $K$ instead of $B$.

To understand the semantics of $\NALSAYS$, first observe the following.  
Suppose that, for all worlds $w'$, it holds that $w \leq w'$ implies $w = w'$.%
\footnote{This condition corresponds to the axiom of excluded middle, hence its imposition creates a classical variant of FOCAL. 
So it makes sense that adding the frame condition would result in the classical semantics of $\Box$.}  
Then the semantics of $\NALSAYS$ simplifies to the standard semantics of $\Box$ in classical modal logic~\cite{HughesC96}:
\begin{multline*}
K,w,v \models \nalSays{\tau}{\phi} \\
\quad\text{iff}\quad \text{for all~} w'' : w \leq_{\mu(\tau)} w'' \text{~implies~} K,w,v \models \phi.
\end{multline*}
That is, a principal believes a formula holds whenever that formula holds in all accessible worlds.
The purpose of the quantification over $w'$, where $w \leq w'$, in the unsimplified semantics of $\NALSAYS$ is to achieve \emph{monotonicity} of the constructive reasoner:
\begin{propo}\label{thm:monotonicity}
If $K,w,v \models \phi$ and $w \leq w'$ then \linebreak $K,w',v \models \phi$.
\end{propo}
\proofinappendix
\noindent That is, whenever $\phi$ holds at a world $w$, if the constructive reasoner is able to reach an extended state of knowledge at world $w'$, then $\phi$ should continue to hold at $w'$.  
Without the quantification over $w'$ in the semantics of $\NALSAYS$, monotonicity is not guaranteed to hold.
Constructive modal logics have, unsurprisingly, also used this semantics for $\Box$~\cite{Simpson94,Wijesekera90},
and a similar semantics has been used in authorization logic~\cite{Garg08}.

Note that, if there do not exist any worlds $w'$ and $w''$ such that $w \leq w' \leq_{\mu(\tau)} w''$, then at $w$, principal $\tau$ will say any formula $\phi$, including $\nalFalse$.  
When a principal says $\nalFalse$ at world $w$, we deem that principal \emph{compromised} at $w$.

As for the semantics of $\NALSPEAKSFOR$, it might be tempting to try defining it as syntactic sugar:  
\begin{equation*}
\nalSpeaksfor{\tau_1}{\tau_2} \quad\equiv\quad
\forall \phi : \nalSays{\tau_1}{\phi} \nalImplies \nalSays{\tau_2}{\phi}
\end{equation*}
However, the formula on the right-hand side is not a well-formed formula of FOCAL, because it quantifies over syntactic formulas. 
So the semantics of $\NALSPEAKSFOR$ cannot interpret it directly in terms of $\NALSAYS$.%
\footnote{It is possible~\cite{GargA08,SchneiderWS11} to instead use second-order quantifiers to achieve a direct interpretation. 
That solution would unnecessarily complicate our semantics by introducing second-order quantifiers solely for the sake of defining $\NALSPEAKSFOR$.}

Instead, the FOCAL semantics of $\NALSPEAKSFOR$ generalizes the classical Kripke semantics of $\NALSPEAKSFOR$~\cite{AbadiBLP93,Howell00}.  
Classically,
\begin{equation}\label{eq:simplsf}
K,w,v \models \nalSpeaksfor{\tau_1}{\tau_2} 
\quad\text{iff}\quad
\mathord{\leq_{\mu(\tau_1)}} \supseteq \mathord{\leq_{\mu(\tau_2)}}.
\end{equation}
\noindent That is, the accessibility relation of $\tau_1$ must be a superset of the accessibility relation of $\tau_2$.  
However, that definition does not account for constructive accessibility, and it even turns out to interact badly with hand-off.

We therefore relax the classical semantics of $\NALSPEAKSFOR$:
\begin{equation}\label{eq:realsf}
K,w,v \models \nalSpeaksfor{\tau_1}{\tau_2}
\quad\text{iff}\quad
 \RA{\tau_1}{w}
   \supseteq \RA{\tau_2}{w} 
\end{equation}
\noindent 
where $\RA{p}{w}$ is defined to be $\mathord{\leq_p}\,|_{[w]_p}$,%
\footnote{%
If $R$ is a binary relation on set $A$, then $R|_{X}$ is the \emph{restriction} of $R$ to $A$, where $X \subseteq A$.
That is, $R|_{X} = \setdef{(x,x')}{(x,x') \in R \text{~and~} x \in X \text{~and~} x' \in X}$.}
and $[w]_p$ is defined to be the set of worlds $w'$ such that $w'$ is reachable from $w$, or vice-versa, by relation $(\mathord{\leq} \cup \mathord{\leq_p})^*$. 
Note that whenever $[w]_p$ equals $W$ (as it would in classical logic\footnote{When frame condition $\mathord{\leq} = W \times W$ is imposed, constructive logic collapses to classical.  Under that condition, every world $w'$ would be reachable from $w$, hence $[w]_p = W$.}), it holds that $\RA{p}{w}$ equals $\leq_p$.

The validity judgment for FOCAL is therefore quite standard, except for $\NALSPEAKSFOR$, where it generalizes classical logic.  
Although we would prefer to adopt a well-known constructive semantics of $\NALSPEAKSFOR$, neither of the two we're aware of seems to work for FOCAL:
ICL~\cite{GargA08} would impose an axiom called Unit that we do not want to include (cf.\ \S\ref{sec:focalevsnal}), 
and {\blsf}~\cite{GenoveseGR12} does not include hand-off~\eqref{eq:handoff}, which we want to optionally support (cf.\ \S\ref{sec:beliefsemanticvalidity} and \S\ref{sec:frameconditions}).

\subsection{Frame conditions}
\label{sec:frameconditions}

\ifframecond
\tikzset{
    world/.style={circle,draw,minimum size=6.5mm,inner sep=2pt},
    edge/.style={-stealth,thick},
}

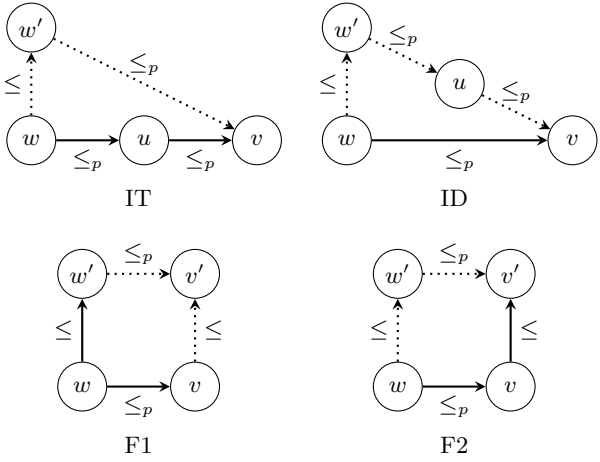
\begin{figure}
\begin{tabular}{cc}
   \begin{tikzpicture}[scale=0.75]
     \node[world] (w) at (0,0) {$w$};
     \node[world] (wp) at (0,2) {$w'$};
     \node[world] (u) at (2,0) {$u$};
     \node[world] (v) at (4,0) {$v$};
     \draw[edge] (w) to node[below]{$\leq_p$} (u);
     \draw[edge] (u) to node[below]{$\leq_p$} (v);
	 \draw[edge,dotted] (w) to node[left]{$\leq$} (wp);
	 \draw[edge,dotted] (wp) to node[above]{$\leq_p$} (v);
   \end{tikzpicture}
&
   \begin{tikzpicture}[scale=0.75]
     \node[world] (w) at (0,0) {$w$};
     \node[world] (wp) at (0,2) {$w'$};
     \node[world] (u) at (2,1) {$u$};
     \node[world] (v) at (4,0) {$v$};
     \draw[edge] (w) to node[below]{$\leq_p$} (v);
	 \draw[edge,dotted] (w) to node[left]{$\leq$} (wp);     
     \draw[edge,dotted] (wp) to node[above]{$\leq_p$} (u);
	 \draw[edge,dotted] (u) to node[above]{$\leq_p$} (v);
   \end{tikzpicture}
\\
IT & ID
\\[2ex]
   \begin{tikzpicture}[scale=0.75]
     \node[world] (w) at (0,0) {$w$};
     \node[world] (wp) at (0,2) {$w'$};
     \node[world] (v) at (2,0) {$v$};
     \node[world] (vp) at (2,2) {$v'$};
     \draw[edge] (w) to node[below]{$\leq_p$} (v);
     \draw[edge] (w) to node[left]{$\leq$} (wp);
	 \draw[edge,dotted] (wp) to node[above]{$\leq_p$} (vp);
	 \draw[edge,dotted] (v) to node[right]{$\leq$} (vp);
   \end{tikzpicture}
&
   \begin{tikzpicture}[scale=0.75]
     \node[world] (w) at (0,0) {$w$};
     \node[world] (wp) at (0,2) {$w'$};
     \node[world] (v) at (2,0) {$v$};
     \node[world] (vp) at (2,2) {$v'$};
     \draw[edge] (w) to node[below]{$\leq_p$} (v);
     \draw[edge,dotted] (w) to node[left]{$\leq$} (wp);
	 \draw[edge,dotted] (wp) to node[above]{$\leq_p$} (vp);
	 \draw[edge] (v) to node[right]{$\leq$} (vp);
   \end{tikzpicture}
\\
F1 & F2
\end{tabular}
\caption{Frame conditions for Kripke semantics\label{fig:frameconditions}}
\end{figure}
\fi

We now return to the discussion begun in \S\ref{sec:modalmodels} of the frame conditions for FOCAL.  
The first two frame conditions we impose help to ensure Says Transparency:
\begin{mrcdefn}{IT}
If $w \leq_p u \leq_p v$, then there exists a $w'$ such that \\ $w \leq w' \leq_p v$.
\end{mrcdefn}
\begin{mrcdefn}{ID} 
If $w \leq_p v$, then there exists a $w'$ and $u$ such that \\ $w \leq w' \leq_p u \leq_p v$.
\end{mrcdefn}
\ifframecond
Figure~\ref{fig:frameconditions} depicts these conditions; dotted lines indicate existentially quantified edges.
\fi
IT helps to guarantee if $\nalSays{p}{\phi}$ then $\nalSays{p}{(\nalSays{p}{\phi})}$; 
ID does the converse.%
\footnote{%
IT and ID are abbreviations for intuitionistic transitivity and intuitionistic density.}

Note how, if $w = w'$, the conditions reduce to the classical definitions of transitivity and density.  
Those classical conditions are exactly what guarantee transparency in classical modal logic.

IT and ID are not quite sufficient to yield transparency.  
By also imposing the following frame condition, we do achieve transparency:%
\footnote{%
F2 is the name given this condition by Simpson~\cite{Simpson94}.}
\begin{mrcdefn}{F2}
If $w \leq_p v \leq v'$, then there exists a $w'$ such that \\ $w \leq w' \leq_p v'.$
\end{mrcdefn}
\ifframecond
F2 is depicted in figure~\ref{fig:frameconditions}.
\fi
It is difficult to motivate F2 solely in terms of authorization logic, though it has been proposed in several Kripke semantics for constructive modal logics~\cite{Simpson94,PlotkinS86,FischerServi81,Ewald86}.
But there are two reasons why F2 is desirable for FOCAL:
\begin{itemize}
\item Assuming F2 holds, IT and ID are not only sufficient but also necessary conditions for transparency---a result that follows from work by Plotkin and Stirling~\cite{PlotkinS86}.  
So in the presence of F2, transparency in FOCAL is precisely characterized by IT and ID.
\item Suppose FOCAL were to be extended with a $\Diamond$ modality.  It could be written $\nalSuspects{\tau}{\phi}$, with semantics $K,w,v \models \nalSuspects{\tau}{\phi}$ iff there exists $w'$ such that $w \leq_{\mu(\tau)} w'$ and $K,w',v \models \phi$.
We would want $\NALSAYS$ and $\NALSUSPECTS$ to interact smoothly.  
For example, it would be reasonable to expect that $\nalNot(\nalSuspects{\tau}{\phi})$ implies $\nalSays{\tau}{\nalNot\phi}$.  For if $\tau$ does not suspect $\phi$ holds anywhere, then $\tau$ should believe $\nalNot\phi$ holds.
Condition F2 guarantees that implication~\cite{PlotkinS86}.
So F2 prepares FOCAL for future extension with a $\NALSUSPECTS$ modality.%
\footnote{%
Were $\NALSUSPECTS$ to be added to FOCAL, it would also be desirable to impose a fourth frame condition:  
if $w \leq w'$ and $w \leq_p v$, then there exists a $v'$ such that $v \leq v'$ and $w' \leq_p v'$.
This condition, named F1 by Simpson~\cite{Simpson94}, guarantees~\cite{PlotkinS86} that $\nalSuspects{\tau}{\phi}$ implies $\nalNot(\nalSays{\tau}{\nalNot\phi})$.
It also guarantees monotonicity (cf.\ proposition~\ref{thm:monotonicity}) for $\NALSUSPECTS$.
\ifframecond Figure~\ref{fig:frameconditions} depicts F1. \fi
Simpson~\cite[p.\ 51]{Simpson94} argues that F1 and F2 could be seen as fundamental, not artificial, frame conditions for constructive modal logics.
}
\end{itemize}

To ensure the validity of hand-off, we impose the following frame condition:
\begin{mrcdefn}{H}
For all principals $p$ and worlds $w$, if there do not exist any worlds $w'$ and $w''$ such that $w \leq w' \leq_p w''$, then, for all $p'$, it must hold that $\RA{p}{w} \subseteq \RA{p'}{w}$.
\end{mrcdefn}
This condition guarantees that if a principal $p$ becomes compromised at world $w$, then the reachable component of its accessibility relation will be a subset of all other principals'.
By the FOCAL semantics of $\NALSPEAKSFOR$, all other principals therefore speak for $p$ at $w$.

Each frame condition above was imposed, not for ad hoc purposes, but because of a specific need in the proof of the soundness result of \S\ref{sec:proofsystem}.
So with appropriate deletion of rules from the proof system, each of the above frame conditions could be eliminated. 
IT and ID should be removed if rules \ruleName{says-li} and \ruleName{says-ri} (from figure~\ref{fig:focalproofsystem}) are removed;
F2 should be removed if rule \ruleName{says-lri} is removed;
and H should be removed if rule \ruleName{sf-i} is removed.

Finally, we impose one additional condition to achieve the equivalence results (theorem~\ref{thm:ktobsound} and proposition~\ref{thm:ktobwf}) of \S\ref{sec:semanticequiv}:
\begin{mrcdefn}{WSF}
$K,w,v \models \nalSpeaksfor{\tau}{\tau'}$ iff, for all $\phi$, if $K,w,v \models \nalSays{\tau}{\phi}$ then $K,w,v \models \nalSays{\tau'}{\phi}$.
\end{mrcdefn}
This condition restricts the class of Kripke models to those where $\NALSPEAKSFOR$ is the \emph{weak speaksfor} connective~\cite{AbadiBLP93,Howell00}.
In fact, we'd prefer to use WSF directly as the semantics of $\NALSPEAKSFOR$ in figure~\ref{fig:focalkripkesemantics}.\footnote{%
If FOCAL included second-order quantification as a logical connective, $\NALSPEAKSFOR$ could be defined as syntactic sugar~\cite{Abadi07}, avoiding the awkwardness of WSF.}
But it wouldn't be a well-founded definition of $\models$, because $\phi$ could itself be $\nalSpeaksfor{\tau}{\tau'}$, leading to a circularity in the semantic definition. 
So we instead impose WSF as a separate axiom.

\section{Semantic Transformation}
\label{sec:semanticequiv}

We have now given two semantics for FOCAL, a belief semantics (\S\ref{sec:beliefsemantics}) and a Kripke semantics (\S\ref{sec:kripkesemantics}).  
How are these two semantics related?
It turns out that a Kripke model can be transformed into a belief model, but the converse does not hold---as we now explain.

\label{sec:ktob}

Given a modal model $K$, there is a natural way to construct a belief model from it:  
assign each principal a worldview containing exactly the formulas that the principal says in $K$.
Call this construction $\ktob$, and let $\ktob(K)$ denote the resulting belief model.

To give a precise definition of $\ktob$, we need to introduce a new notation.  
Given a principal $p \in P$, formula $\nalSays{p}{\phi}$ is not necessarily well-formed, because $p$ is not necessarily a syntactic term.  
So let $K,w,v \models \nalSays{\hat{p}}{\phi}$ be defined as follows: 
for all $w'$ and $w''$ such that $w \leq w' \leq_p w''$, it holds that $K,w'',v \models \phi$.
This definition simply unrolls the semantics of $\NALSAYS$ to produce something well-formed.%
\footnote{Another solution would be to stipulate that every principal $p$ can be named by a term $\hat{p}$ in the syntax.}

The precise definition of $\ktob$ is as follows:
if $K = (W, \mathord{\leq}, s,$ $P, A)$, then $\ktob(K)$ is belief model $(W,\leq,s,P,\worldview)$, where $\worldview(w,p,v)$ is defined to be $\setdef{\phi}{K,w,v \models \nalSays{\hat{p}}{\phi}}$.   

Our first concern is whether $\ktob(K)$ produces a belief model that is equivalent to $K$.  
In particular, a formula should be valid in $K$ iff it is valid in $\ktob(K)$.  
Construction $\ktob$ does produce equivalent models:
\begin{theorem}\label{thm:ktobsound}
For all $K$, $w$, $v$, and $\phi$, 
$K,w,v \models \phi$ iff $\ktob(K),w,v \models \phi$.
\end{theorem}

Our second concern is whether $\ktob(K)$ satisfies all the conditions required by \S\ref{sec:beliefsemantics}:  Worldview Monotonicity, Worldview Equality, Worldview Closure, Says Transparency, and Belief Hand-off.  
If a belief model $B$ does satisfy these conditions, then $B$ is \emph{well-formed}. 
And modal model $K$ is well-formed if it satisfies all the conditions required by \S\ref{sec:kripkesemantics}: Accessibility Equality, IT, ID, F2, H, and WSF.
Construction $\ktob$ does, indeed, produce well-formed belief models:
\begin{propo}\label{thm:ktobwf}
For all well-formed modal models $K$, belief model $\ktob(K)$ is well-formed.
\end{propo}
\proofinappendix

\begin{figure*}
\begin{equation*}
\renewcommand{\arraystretch}{3}
\begin{array}{c}
\infer[\smallRuleName{hyp}]{\Gamma,\phi \proves \phi}{}
\quad
\infer[\smallRuleName{weak}]{\Gamma,\psi \proves \phi}{\Gamma \proves \phi}
\quad
\infer[\smallRuleName{true-i}]{\Gamma \proves \nalTrue}{}
\quad
\infer[\smallRuleName{false-e}]{\Gamma \proves \phi}{\Gamma \proves \nalFalse}
\quad
\infer[\smallRuleName{and-i}]{\Gamma \proves \phi \nalAnd \psi}{\Gamma \proves \phi & \Gamma \proves \psi}
\quad
\infer[\smallRuleName{and-le}]{\Gamma \proves \phi}{\Gamma \proves \phi \nalAnd \psi}
\\
\infer[\smallRuleName{and-re}]{\Gamma \proves \psi}{\Gamma \proves \phi \nalAnd \psi}
\quad
\infer[\smallRuleName{or-li}]{\Gamma \proves \phi_1 \nalOr \phi_2}{\Gamma \proves \phi_1}
\quad
\infer[\smallRuleName{or-ri}]{\Gamma \proves \phi_1 \nalOr \phi_2}{\Gamma \proves \phi_2}
\quad
\infer[\smallRuleName{or-e}]{\Gamma \proves \psi}{\Gamma \proves \phi_1 \nalOr \phi_2 & \Gamma,\phi_1 \proves \psi & \Gamma,\phi_2 \proves \psi}
\quad
\infer[\smallRuleName{imp-i}]{\Gamma \proves \phi \nalImplies \psi}{ \Gamma,\phi \proves \psi}
\\ 
\infer[\smallRuleName{imp-e}]{\Gamma \proves \psi}{\Gamma \proves \phi & \Gamma \proves \phi \nalImplies \psi}
\quad
\infer[\smallRuleName{not-i}]{\Gamma \proves \nalNot\phi}{\Gamma,\phi \proves \nalFalse}
\quad
\infer[\smallRuleName{not-e}]{\Gamma \proves \nalFalse}{\Gamma \proves \phi & \Gamma \proves \nalNot\phi}
\quad
\infer[\smallRuleName{forall-i}]{\Gamma \proves \nalForall{x}{\phi}}{\Gamma \proves \phi & x \not\in \FV(\Gamma)}
\quad
\infer[\smallRuleName{forall-e}]{\Gamma \proves \phi\subst{\tau}{x}}{\Gamma \proves \nalForall{x}{\phi}}
\\
\infer[\smallRuleName{exists-i}]{\Gamma \proves \nalExists{x}{\phi}}{\Gamma \proves \phi\subst{\tau}{x}}
\quad
\infer[\smallRuleName{exists-e}]{\Gamma \proves \psi}{\Gamma \proves \nalExists{x}{\phi} & \Gamma,\phi \proves \psi & x \not\in \FV(\Gamma,\psi)}
\quad
\infer[\smallRuleName{eq-r}]{\Gamma \proves \tau = \tau}{}
\quad
\infer[\smallRuleName{eq-s}]{\Gamma \proves \tau_2 = \tau_1}{\Gamma \proves \tau_1 = \tau_2}
\\
\infer[\smallRuleName{eq-t}]{\Gamma \proves \tau_1 = \tau_3}{\Gamma \proves \tau_1 = \tau_2 & \Gamma \proves \tau_2 = \tau_3}
\quad
\infer[\smallRuleName{eq-fun}]{\Gamma \proves f(\tau_1,\ldots,\tau_n) = f(\tau'_1,\ldots,\tau'_n)}{\Gamma \proves \tau_i = \tau'_i}
\quad
\infer[\smallRuleName{eq-rel}]{\Gamma \proves r(\tau'_1,\ldots,\tau'_n)}{\Gamma\proves r(\tau_1,\ldots,\tau_n) & \Gamma \proves \tau_i = \tau'_i}
\\
\infer[\smallRuleName{says-lri}]{\nalSays{\tau}{\Gamma} \proves \nalSays{\tau}{\phi}}{\Gamma \proves \phi}
\quad
\infer[\smallRuleName{says-li}]{\nalSays{\tau}{\Gamma} \proves \nalSays{\tau}{\phi}}{\Gamma \proves \nalSays{\tau}{\phi}}
\quad
\infer[\smallRuleName{says-ri}]{\nalSays{\tau}{\Gamma} \proves \nalSays{\tau}{\phi}}{\nalSays{\tau}{\Gamma} \proves \phi}
\quad
\infer[\smallRuleName{sf-i}]{\Gamma \proves \nalSpeaksfor{\tau_1}{\tau_2}}{\Gamma \proves \nalSays{\tau_2}{(\nalSpeaksfor{\tau_1}{\tau_2})}}
\\
\infer[\smallRuleName{sf-e}]{\Gamma \proves \nalSays{\tau_2}{\phi}}{\Gamma \proves \nalSpeaksfor{\tau_1}{\tau_2} & \Gamma \proves \nalSays{\tau_1}{\phi}}
\quad
\infer[\smallRuleName{sf-r}]{\Gamma \proves \nalSpeaksfor{\tau}{\tau}}{}
\quad
\infer[\smallRuleName{sf-t}]{\Gamma \proves \nalSpeaksfor{\tau_1}{\tau_3}}{\Gamma \proves \nalSpeaksfor{\tau_1}{\tau_2} & \Gamma \proves \nalSpeaksfor{\tau_2}{\tau_3}}
\end{array}
\end{equation*}
\caption{FOCAL derivability judgment\label{fig:focalproofsystem}}
\end{figure*}

We might wonder whether there is a construction that can soundly transform belief models into Kripke models.
Consider trying to transform the following belief model $B$ into a Kripke model:
\begin{quote}
$B$ has a single world $w$ and a proposition (i.e., a nullary relation) $X$, such that, for all $v$, it holds that $B,w,v \not\models X$.
Suppose that principal $p$'s worldview contains $X$---i.e., for all $v$, it holds that $X \in \worldview(w,p,v)$---and that $p$'s worldview does not contain $\nalFalse$.
By the semantics of $\NALSAYS$, it holds that $B,w,v \models \nalSays{p}{X}$.
\end{quote}
When transforming $B$ to a Kripke model $K$, what edges could we put in $\leq_p$?
There are only two choices:  $\leq_p$ could be empty, or $\leq_p$ could contain the single edge $(w,w)$.
If $\leq_p$ is empty, then $p$ is compromised, hence $p$ says $\nalFalse$.  
That contradicts our assumption that $\nalFalse$ is not in $p$'s worldview.
If $w \leq_p w$, then for $w'$ and $w''$ such that $w \leq w' \leq_p w''$, it does not hold that $K,w'',v \models X$, because $w$ and $w''$ can only be instantiated as $w$, and because $B,w,v \not\models X$.  
Hence $p$ does not say $X$.  
That contradicts our assumption that $X$ is in $p$'s worldview.
So we cannot construct an accessibility relation $\leq_p$ that causes the resulting Kripke semantics to preserve validity of formulas from the belief semantics.

There is, therefore, no construction that can soundly transform belief models into Kripke models---unless, perhaps, the set of worlds is permitted to change.
We conjecture that it is possible to synthesize a new set of possible worlds, and equivalence relations on them, yielding a Kripke model that preserves validity of formulas from the belief model.

%
%
%

\section{Proof System}
\label{sec:proofsystem}

FOCAL's derivability judgment is written $\Gamma \proves \phi$ where $\Gamma$ is a set of formulas called the \emph{context}.%
\footnote{%
These formulas are \emph{localized hypotheses}, which the proof system uses instead of the hypothetical judgments found in natural deduction systems.
Similar to the left-hand side $\Gamma$ of a sequent $\Gamma \Longrightarrow \Delta$, the localized hypotheses are assumptions being used to derive right-hand side $\Delta$.
Unlike a sequent, $\Gamma$ is a set, not a sequence.}
As is standard, we write $\proves\phi$ when $\Gamma$ is the empty set.  
In that case, $\phi$ is a \emph{theorem}. 
We write $\Gamma,\phi$ to denote $\Gamma \cup \{\phi\}$.

Figure~\ref{fig:focalproofsystem} presents the proof system.
In it, $\phi\subst{\tau}{x}$ denotes capture-avoiding substitution of $\tau$ for $x$ in $\phi$.
The first-order fragment of the proof system is routine (e.g.,~\cite{SorensenU06,vanDalen04,NegrivP01}).%
\footnote{Under the usual constructive definition of $\nalNot \phi$ as $\phi \nalImplies \nalFalse$, rules \ruleName{not-i} and \ruleName{not-e} are merely admissible rules and could be eliminated from the proof system.}
Because of \ruleName{imp-i}, the deduction theorem holds for FOCAL~\cite{HakliN12}.
\ruleName{says-lri}, \ruleName{says-li}, and \ruleName{says-ri} use notation $\nalSays{\tau}{\Gamma}$, which means that $\tau$ says all the formulas in set $\Gamma$.  
Formally, $\nalSays{\tau}{\Gamma}$ is defined as $\setdef{\nalSays{\tau}{\phi}}{\phi \in \Gamma}$.

\ruleName{says-lri} corresponds~\cite{HughesC96} to standard axiom $K$ along with rule $N$ from epistemic logic; \ruleName{says-ri}, to standard axiom $4$; and  \ruleName{says-li}, to the converse $C4$~\cite{Abadi08,Chellas80} of $4$: 
\begin{align*}
K:\quad &\proves (\nalSays{p}{(\phi \nalImplies \psi)}) \nalImplies (\nalSays{p}{\phi}) \nalImplies (\nalSays{p}{\psi}), \\
N:\quad &\text{From $\proves \phi$ infer $\proves \nalSays{p}{\phi}$}, \\
4:\quad &\proves (\nalSays{p}{\phi}) \nalImplies (\nalSays{p}{(\nalSays{p}{\phi})}), \\
C4:\quad &\proves (\nalSays{p}{(\nalSays{p}{\phi})}) \nalImplies (\nalSays{p}{\phi}).
\end{align*}
$K$ and \ruleName{says-lri} mean that \emph{modus ponens} applies inside $\NALSAYS$.  
They correspond to Worldview Closure.
$C4$ and $4$, along with \ruleName{says-li} and \ruleName{says-ri}, mean that $\nalSays{p}{(\nalSays{p}{\phi})}$ is equivalent to $\nalSays{p}{\phi}$; they correspond to Says Transparency in the belief semantics.
In the Kripke semantics, \ruleName{says-ri} corresponds to IT; and \ruleName{says-li}, to ID.
By including rules corresponding to $4$ and $C4$, it is not our intent to argue that those axioms are necessary in authorization logics (which is debatable); rather, our intent is just to show how to support them.

\ruleName{sf-i} corresponds to hand-off~\eqref{eq:handoff}.  
\ruleName{sf-e} uses $\NALSPEAKSFOR$ to deduce beliefs.  
\ruleName{sf-r} and \ruleName{sf-t} state that $\NALSPEAKSFOR$ is reflexive and transitive.

The usual sequent calculus structural rules of contraction and exchange are admissible.
But weakening (our rule \ruleName{weak}) is not admissible:
it must be directly included in the proof system, because the conclusions of $\ruleName{says-\{lri,li,ri\}}$ capture their entire context $\Gamma$ inside $\NALSAYS$.

\subsection{Soundness}

Our first soundness theorem for FOCAL states that if $\phi$ is provable from assumptions $\Gamma$, and that if a belief model validates all the formulas in $\Gamma$, then that model must also validate $\phi$.  
Therefore, any provable formula is valid in the belief semantics:
\begin{theorem}\label{thm:beliefsound}
If $\Gamma \proves \phi$ and $B,w,v \models \Gamma$, then $B,w,v \models \phi$.
\end{theorem}
\proofinappendix
\noindent 
We have mechanized the proof of this theorem in Coq.  
The result is, to our knowledge, the first proof of soundness for an authorization logic w.r.t.\ a belief semantics.
The proof of theorem \ref{thm:beliefsound} relies on the following proposition, which states monotonicity of validity w.r.t.\ $\leq$:
\begin{propo}\label{thm:beliefmonotonicity}
If $B,w,v \models \phi$ and $w \leq w'$ then \linebreak $B,w',v \models \phi$.
\end{propo}
The proof of it is also mechanized in Coq.

Our second soundness theorem for FOCAL states that any provable formula is valid in the Kripke semantics:
\begin{theorem}\label{thm:kripkesound}
If $\Gamma \proves \phi$ and $K,w,v \models \Gamma$, then $K,w,v \models \phi$.
\end{theorem}
\proofinappendix
\noindent The proof of that theorem relies on proposition~\ref{thm:monotonicity} (monotonicity of the Kripke semantics).
We also have mechanized the proofs of theorem \ref{thm:kripkesound} and proposition~\ref{thm:monotonicity} in Coq.  

\subsection{State in distributed systems}

\label{sec:focalevsnal}
FOCAL was derived from CDD~\cite{Abadi07} and NAL~\cite{SchneiderWS11}.
But we deliberately designed the FOCAL proof system such that its theory differs in one important way from theirs.
We discuss our motivation for this change, next.

There are two standard ways of ``importing'' beliefs into a principal's worldview.  
The first is rule $N$ from \S\ref{sec:proofsystem}, also known as the rule of Necessitation:  
from $\proves \phi$, infer $\proves \nalSays{p}{\phi}$.
The second is an axiom known as Unit:
$\proves \phi \nalImplies (\nalSays{p}{\phi})$.
Though superficially similar, it is well-known that Necessitation and Unit lead to different theories.  
Abadi~\cite{Abadi08} explores some of the proof-theoretic differences, particularly some of the surprising consequences of Unit in classical authorization logic.  
In the example below, we focus on one difference that does not seem to have been explored in constructive authorization logic:

\begin{example}
Machines $M_1$ and $M_2$ execute processes $P_1$ and $P_2$, respectively.   
$M_1$ has a register $R$.  
Let $Z$ be a proposition representing ``register $R$ is currently set to zero.''  
According to Unit, $\proves Z \nalImplies (\nalSays{P_1}{Z})$ and $\proves Z \nalImplies (\nalSays{P_2}{Z})$.  
The former means that a process on a machine knows the current contents of a register on that machine; the latter means that a process on a different
machine must also know the current contents of the register. 
But according to Necessitation, if $\proves Z$ then $\proves \nalSays{P_1}{Z}$ and $\proves \nalSays{P_2}{Z}$.  
Only if $R$ is guaranteed to be constant---i.e., it can never at any time be anything other than zero---must the two processes say so.  
\end{example}

Unit, therefore, is appropriate when propositions (or relations or functions)
represent global state upon which all principals are guaranteed to agree.
But when propositions represent local state that could be unknown to some principals, Unit would arguably be an invalid axiom.
A countermodel demonstrating Unit's invalidity is easy to construct---for example, stipulate a world $w$ at which $Z$ holds, and let $P_1$'s worldview contain $Z$ but $P_2$'s worldview not contain $Z$.
That countermodel doesn't apply to Necessitation, because $Z$ is not a theorem in it, therefore the principals may disagree on $Z$'s validity.

Prior work has objected to Unit for other reasons (cf.\ \S\ref{sec:relatedwork}), but not for this difference between local and global state.
We are unaware of any authorization logic that rejects Necessitation, which is widely accepted along with axiom $K$ (cf.\ \S\ref{sec:proofsystem}) in \emph{normal modal logic}~\cite{HughesC96}.  

FOCAL is designed for reasoning about state in distributed systems, where principals (such as machines) may have local state, and where global state does not necessarily exist---the reading at a clock, for example, is not agreed upon by all principals. 
So Unit would be invalid for FOCAL principals;  Necessitation is the appropriate choice.
We therefore include Necessitation in FOCAL in the form of rule \ruleName{says-lri}.
Having that rule in our proof system is equivalent to having both Necessitation and $K$ in a natural-deduction proof system~\cite[p.\ 214, where \ruleName{says-lri} is called \ruleName{lr}]{HughesC96}.
Unit, on the other hand, is invalid in FOCAL's semantics, and FOCAL's proof system is sound w.r.t.\ its semantics, so it's impossible to derive Unit in FOCAL.

Similarly, NAL principals do not necessarily agree upon global state.  
NAL does include Necessitation as an inference rule and does not include Unit as an axiom.
However, NAL permits Unit to be derived as a theorem:%
\footnote{%
Rules \ruleName{nal-imp-i} and \ruleName{nal-says-i} are given by Schneider et al.~\cite{SchneiderWS11}.
The brackets around $\phi$ at the top of the proof tree indicate that it is used as a hypothesis~\cite{vanDalen04}.
The appearance of ``1'' as a super- and subscript indicate where the hypothesis is introduced and cancelled.
}
\begin{equation*}
\infer[\smallRuleName{nal-imp-i$_1$}]
{\phi \nalImplies \nalSays{p}{\phi}}
{
  \infer[\smallRuleName{nal-says-i}]
  {\nalSays{p}{\phi}}
  {[\phi]^1}
}
\end{equation*}
NAL's proof system is, therefore, arguably unsound w.r.t.\ our belief semantics:
there is a formula (Unit) that is a theorem of the system but that is not semantically valid. 

\label{sec:focalevscdd}
NAL extends CDD's proof system, so we might suspect that CDD is also unsound w.r.t.\ our semantics.  
And it is.
However, CDD has been proved sound w.r.t.\ a \emph{lax logic} semantics~\cite{GargA08}.
That semantics employs a different intuition about $\NALSAYS$ than NAL.
CDD understands $\nalSays{p}{\phi}$ to mean ``when combining the [statement $\phi$] that the [guard] believes with those that [$p$] contributes, the [guard] can conclude $\phi${\ldots}the [guard's] participation is left implicit''~\cite[p.\ 13]{Abadi07}.
In other words, the guard's beliefs are imported into $p$'s beliefs at each world.
That results in a different meaning of $\NALSAYS$ than FOCAL or NAL employs.

Since Abadi's invention of CDD~\cite{Abadi07}, the $\NALSAYS$ connective is frequently assumed to satisfy the \emph{monad}~\cite{Moggi91} laws, which include Unit.
But FOCAL rejects Unit, so FOCAL's $\NALSAYS$ connective is not a monad.  
The monad laws also include an axiom named Bind, which turns out to be invalid in FOCAL's semantics.%
\footnote{The terms ``monad,'' ``lax logic,'' and the combination of axioms Unit and Bind all three convey the same mathematical structure, so it's not surprising that FOCAL differs from all of them.}
We don't know whether rejecting the monad laws will have any practical impact on FOCAL. 
But the seminal authorization logic, ABLP~\cite{AbadiBLP93}, didn't adopt the monad laws.
Likewise, Garg and Pfenning~\cite{GargP10} reject Unit in their authorization logic BL$_0$;
they demonstrate that Unit leads to counterintuitive interpretations of some formulas involving delegation.  
And Abadi~\cite{Abadi03} notes that Unit ``should be used with caution (if at all),'' suggesting that it be replaced with the weaker axiom $(\nalSays{p}{\phi}) \nalImplies (\nalSays{q}{\nalSays{p}{\phi})}$.
Genovese et al.~\cite{GenoveseGR12} carry out that suggestion.
So in rejecting the monad laws, FOCAL is at least in good company.

\section{Related Work}
\label{sec:relatedwork}

FOCAL has the first formal belief semantics of any authorization logic.  
To our knowledge, belief semantics have been used in only one other authorization logic, and that logic---NAL~\cite{SchneiderWS11}---has only an informal semantics based on worldviews.

But many of the pieces of FOCAL, including its semantics and proof system, are naturally derived from previous work.  
We summarize here what we borrowed vs.\ what we invented; the main body of the paper contains detailed citations.
FOCAL's belief semantics is a standard first-order constructive semantics, but the addition of worldviews to interpret $\NALSAYS$ and $\NALSPEAKSFOR$ is novel (with the exception of NAL, which used worldviews informally).
FOCAL's Kripke semantics for everything except $\NALSPEAKSFOR$ is likewise standard, and its frame conditions (except H and WSF) are already well-known in constructive modal logic, but the application of IT and ID to authorization logic seems to be novel.  
FOCAL's proof system, excluding $\NALSAYS$ and $\NALSPEAKSFOR$, is a straightforward first-order constructive proof system.
The fragment for $\NALSAYS$ is our own adaptation of modal-logic natural-deduction rules for the $\Box$ connective.
The fragment for $\NALSPEAKSFOR$ corresponds to standard definitions used in many authorization logics.

Semantic structures similar to our belief models have been investigated in the context of epistemic logic~\cite{Eberle74, MooreH79, FaginHMV95}.  
Konolige~\cite{Konolige86} proves an equivalence result for classical propositional logic similar to our theorem~\ref{thm:ktobsound}.

Garg and Abadi~\cite{GargA08} give a Kripke semantics for a logic they call ICL, which could be regarded as a propositional fragment of FOCAL. 
The ICL semantics of $\NALSAYS$, however, uses \emph{invisible} worlds to permit principals to be oblivious to the truth of formulas at some worlds.
That makes Unit (\S\ref{sec:focalevsnal}) valid in ICL, whereas Unit is invalid in FOCAL.

Garg~\cite{Garg08} studies the proof theory of a logic called {\dtlz},
and gives a Kripke semantics that uses both invisible worlds and \emph{fallible} worlds, at which $\nalFalse$ is permitted to be valid.
Instead of Unit, it uses the axiom $\nalSays{p}{((\nalSays{p}{\phi}) \nalImplies \phi)}$.
That axiom is unsound in FOCAL.
{\dtlz} does not have a $\NALSPEAKSFOR$ connective.  

Genovese et al.~\cite{GenoveseGR12} study several uses for Kripke semantics with an authorization logic they call {\blsf}, which also could be regarded as a propositional fragment of FOCAL.
They show how to generate evidence for why an access should be denied, how to find all logical consequences of an authorization policy, and how to determine which additional credentials would allow an access.
However, the Kripke semantics of {\blsf} differs from FOCAL's in its interpretation of both $\NALSAYS$ and $\NALSPEAKSFOR$, so the results of Genovese et al.~are not immediately applicable to FOCAL.

Garg and Pfenning~\cite{GargP06} prove \emph{non-interference} properties for a first-order constructive authorization logic. 
Such properties mean that one principal's beliefs cannot interfere with another principal's beliefs unless there is some trust relationship between those principals.
Abadi~\cite{Abadi07} also proves such a property for dependency core calculus (DCC), which is the basis of authorization logic CDD.
We conjecture that similar properties could be proved for FOCAL.

\section{Concluding Remarks}
\label{sec:conclusion}

This work began with the idea of giving a Kripke semantics to NAL.
Proving soundness---at first on paper, not in Coq---turned out to be surprising, because Unit is semantically invalid but derivable in NAL (\S\ref{sec:focalevsnal}).
The complexity of the resulting Kripke semantics motivated us to seek a simpler semantics.
We were inspired by the informal worldview semantics of the NAL rationale~\cite{SchneiderWS11} and elaborated that into our belief semantics (\S\ref{sec:beliefsemantics}).  
In future work, we plan to upgrade FOCAL to handle NAL's advanced features, including \emph{intensional group principals}.

Mechanizing the proofs of soundness in Coq was frequently rewarding.  
It exposed several bugs (in either our proof system or our semantics) and gave us high confidence in the correctness of the result.  
We expect further benefits, too.
Other researchers can now use our formalization as a basis for mechanizing results about authorization logics.  
And from the formalization of the FOCAL proof system in Coq, we could next extract a \emph{verified theorem checker}.  
It would input a proof of a FOCAL formula, expressed in the FOCAL proof system, and output whether the proof is correct.  
Coq would verify that the checker correctly implements the FOCAL proof system.
After FOCAL is upgraded to handle all of NAL's features, the resulting theorem checker could replace the current Nexus~\cite{SirerBRSWWS11} theorem checker, which is implemented in C.
A verified theorem checker would arguably be more trustworthy than the C implementation, thus increasing the trustworthiness of the operating system. 

Our goal was to increase the trustworthiness of authorization logics, hence our concentration on soundness results.
Another worthwhile goal would to be increase the utility of authorization logics, and toward that end we could investigate the \emph{completeness} of FOCAL:
are all valid formulas provable?
A few authorization logics---ICL~\cite{GargA08}, {\dtlz}~\cite{Garg08}, and {\blsf}~\cite{GenoveseGR12}---do have completeness results for Kripke semantics; however, none of those is immediately applicable to FOCAL.%
\footnote{%
ICL uses a lax logic semantics that is incompatible with FOCAL's definition of $\NALSAYS$.  
{\dtlz} uses a Kripke semantics with \emph{invisible} and \emph{fallible} worlds, and it omits the $\NALSPEAKSFOR$ connective.
And {\blsf} encodes $\NALSPEAKSFOR$ as a first-order relation, rather than defining it with accessibility relations, and it does not provide a weak speaksfor semantics. 
}  
We leave adaptation of them as future work. 


\section*{Acknowledgments}

{
Fred B.\ Schneider consulted on the design of the proof system and the Kripke semantics, and he suggested the idea of proving an equivalence between belief and Kripke semantics. 
We thank him, Mart\'{i}n Abadi, Deepak Garg, Joe Halpern, Dexter Kozen, Colin Stirling, and Kevin Walsh for discussions related to this work.  
We also thank Adam Chlipala, Kristopher Micinski, and the coq-club mailing list for assistance with Coq.
The anonymous reviewers of CSF 2013 and CCS 2013 provided invaluable feedback.
The Kripke semantics of FOCAL was created while Clarkson was a postdoctoral researcher at Cornell.
This work was supported in part by AFOSR grants F9550-06-0019, FA9550-11-1-0137, and FA9550-12-1-0334, NSF grants 0430161, 0964409, and CCF-0424422 (TRUST), ONR grants N00014-01-1-0968 and N00014-09-1-0652, and a grant from Microsoft.
}

\pagebreak

\bibliographystyle{abbrv}
\bibliography{../bib/nal}

\def\appendix{\par
\section*{APPENDIX: PROOFS}
\setcounter{section}{0}
 \setcounter{subsection}{0}
 \def\thesection{\Alph{section}} }

\appendix


\medskip 

\subsection*{Proposition \ref{thm:monotonicity}.}

If $K,w,v \models \phi$ and $w \leq w'$ then $K,w',v \models \phi$.

\begin{proof}
By structural induction on $\phi$.  
This proof has been mechanized in Coq.
\end{proof}

\subsection*{Theorem \ref{thm:ktobsound}.}
For all $K$, $w$, $v$, and $\phi$, it holds that 
$K,w,v \models \phi$ iff \linebreak $\ktob(K),w,v \models \phi$.

\begin{proof}  First, we show the forward direction:  
$K,w,v \models \phi$ implies $\ktob(K),w,v \models \phi$.
All of the cases except $\NALSAYS$ and $\NALSPEAKSFOR$ are straightforward, because those are the only two cases where the interpretation of formulas differs in the two semantics.
\begin{itemize}
\item Case $\phi = \nalSays{\tau}{\psi}$.  
Suppose $K,w,v \models \nalSays{\tau}{\psi}$.
By the definition of $\ktob$, formula $\psi \in \worldview(w,\mu(\tau),v)$.
By the belief semantics of $\NALSAYS$, it must hold that $\ktob(K),w,v \models \nalSays{\tau}{\psi}$.
\item Case $\phi = \nalSpeaksfor{\tau}{\tau'}$.
Assume $K,w,v \models \nalSpeaksfor{\tau}{\tau'}$.
We need to show that, for all $w' \geq w$, it holds that $\worldview(w',\mu(\tau),v) \subseteq \worldview(w',\mu(\tau'),v)$.  
So let $w'$ and $\psi$ be arbitrary such that $w' \geq w$ and $\psi \in \worldview(w',\mu(\tau),v)$, and we'll show that $\psi \in \worldview(w',\mu(\tau'),v)$.
By the definition of $\ktob$, it holds that $K,w',v \models \nalSays{\tau}{\psi}$.
Note that, by proposition~\ref{thm:monotonicity} and our original assumption, we have that $K,w',v \models \nalSpeaksfor{\tau}{\tau'}$.
From those last two facts, and from the Kripke semantics of $\NALSAYS$ and $\NALSPEAKSFOR$, it follows that $K,w',v \models \nalSays{\tau'}{\psi}$.
By the definition of $\ktob$, it therefore holds that $\psi \in \worldview(w',\mu(\tau'),v)$.
\end{itemize}

Second, we show the backward direction:
$K,w,v \models \phi$ is implied by $\ktob(K),w,v \models \phi$.
Again, all of the cases except $\NALSAYS$ and $\NALSPEAKSFOR$ are straightforward, because those are the only two cases where the interpretation of formulas differs in the two semantics.
\begin{itemize}
\item Case $\phi = \nalSays{\tau}{\psi}$.
Suppose $\ktob(K),w,v \models \nalSays{\tau}{\psi}$.
By the belief semantics of $\NALSAYS$, we have that $\psi \in \worldview(w,\mu(\tau),v)$.
By the definition of $\ktob$, it holds that $K,w,v \models \nalSays{\tau}{\psi}$.
\item Case $\phi = \nalSpeaksfor{\tau}{\tau'}$.
Assume $\ktob(K),w,v \models \nalSpeaksfor{\tau}{\tau'}$.
By the belief semantics of $\NALSPEAKSFOR$, we have that, for all $w' \geq w$, it holds that $\worldview(w',\mu(\tau),v) \subseteq \worldview(w',\mu(\tau'),v)$.
Let $w'$ be $w$.  Then $\worldview(w,\mu(\tau),v) \subseteq \worldview(w,\mu(\tau'),v)$.
By the definitions of $\ktob$ and subset, it follows that, for all $\phi$, if $K,w,v \models \nalSays{\tau}{\phi}$ then $K,w,v \models \nalSays{\tau'}{\phi}$.
By WSF, we therefore have that $K,w,v \models \nalSpeaksfor{\tau}{\tau'}$.\qedhere
\end{itemize}
\end{proof}

\subsection*{Proposition \ref{thm:ktobwf}.}
For all well-formed modal models $K$, belief model $\ktob(K)$ is well-formed.
\begin{proof}
Let $B = \ktob(K)$.  
For $B$ to be well-formed it must satisfy several conditions, which were defined in \S\ref{sec:beliefsemantics}.  
We now show that these hold for any such $B$ constructed by $\ktob$.
\begin{enumerate}
\item Worldview Monotonicity.  
Assume $w \leq w'$ and $\phi \in \worldview(w,p,v)$.
By the latter assumption and the definition of $\ktob$, we have that $K,w,v \models \nalSays{\hat{p}}{\phi}$.  
From proposition~\ref{thm:monotonicity}, it follows that $K,w',v \models \nalSays{\hat{p}}{\phi}$.  
By the definition of $\ktob$, it then holds that $\phi \in \worldview(w',p,v)$.
Therefore $\worldview(w,p,v) \subseteq \worldview(w',p,v)$.
\item Worldview Equality. 
Assume $p \peq p'$.
Then by Accessibility Equality, $\leq_p$ equals $\leq_{p'}$.  
By the Kripke semantics of $\NALSAYS$, it follows that $K,w,v \models \nalSays{p}{\phi}$ iff  $K,w,v \models \nalSays{p'}{\phi}$.  
By the definition of $\ktob$, therefore, $\worldview(w,p,v) = \worldview(w,p',v)$.
\item Worldview Closure. 
Assume $\Gamma \subseteq \worldview(w,p,v)$ and $\Gamma \models \phi$, that is, $\phi$ is a logical consequence of $\Gamma$ in belief structure $B$.
By the definition of $\ktob$, we have $\worldview(w,p,v) = \setdef{\phi}{K, w, v \models \nalSays{\hat{p}}{\phi}}$.
So for all $\psi \in \Gamma$, it holds that $K, w, v \models \nalSays{\hat{p}}{\psi}$.  
By the Kripke semantics of $\NALSAYS$, it follows that for all $w'$ and $w''$ such that $w \leq w' \leq_{p} w''$, it holds that $K,w'',v \models \psi$.  
Thus $K,w'',v \models \Gamma$.
So $\ktob(K),w'',v \models \Gamma$ by theorem \ref{thm:ktobsound}.
By our initial assumption that $\Gamma \models \phi$, it follows that $\ktob(K),w'',v \models \phi$.
Again applying theorem \ref{thm:ktobsound}, we have that $K,w'',v \models \phi$. 
By the Kripke semantics of $\NALSAYS$, it follows that $K,w,v \models \nalSays{\hat{p}}{\phi}$.
Therefore, by the definition of $\ktob$, we have $\phi \in \worldview(w,p,v)$.
\item Says Transparency.
We prove the ``iff'' by proving both directions independently.

($\Rightarrow$)  
Assume $\phi \in \worldview(w,p,v)$.  
By the definition of $\ktob$, it holds that $K,w,v \models \nalSays{\hat{p}}{\phi}$. 
From IT and F2, it follows that $K,w,v \models \nalSays{\hat{p}}{(\nalSays{\hat{p}}{\phi})
}$.
By the definition of $\ktob$, therefore, $(\nalSays{\hat{p}}{\phi}) \in \worldview(w,p,v)$.

($\Leftarrow$)
Assume $(\nalSays{\hat{p}}{\phi}) \in \worldview(w,p,v)$.  
By the definition of $\ktob$, it holds that $K,w,v \models \nalSays{\hat{p}}{(\nalSays{\hat{p}}{\phi})
}$. 
From ID, it follows that $K,w,v \models \nalSays{\hat{p}}{\phi}$.
By the definition of $\ktob$, therefore, $\phi \in \worldview(w,p,v)$.
\item Belief Hand-off.
We actually prove a stronger result---an ``iff'' rather than just an ``if''.
By the definitions of subset and $\ktob$, we have that 
$\worldview(w,p,v) \subseteq  \worldview(w,q,v)$ holds iff
for all $\phi$, if $K,w,v \models \nalSays{\hat{p}}{\phi}$ then 
$K,w,v \models \nalSays{\hat{q}}{\phi}$.
By WSF, that holds iff $K,w,v \models \nalSpeaksfor{\hat{p}}{\hat{q}}$.
By the fact below, that holds iff
$K,w,v \models \nalSays{\hat{q}}{(\nalSpeaksfor{\hat{p}}{\hat{q}})}$.
By the definition of $\ktob$, that holds iff 
$\nalSpeaksfor{\hat{q}}{\hat{p}} \in \worldview(w, q, v)$.

Fact:  in the Kripke semantics,  $\models \nalSays{\hat{q}}{(\nalSpeaksfor{\hat{p}}{\hat{q}})} \linebreak\iff \nalSpeaksfor{\hat{p}}{\hat{q}}$.  
The proof of that fact has been mechanized in Coq.\qedhere
\end{enumerate}
\end{proof}

\subsection*{Theorem \ref{thm:beliefsound}.}
If $\Gamma \proves \phi$ and $B,w,v \models \Gamma$, then $B,w,v \models \phi$.
\begin{proof}
By induction on the derivation of $\Gamma \proves \phi$.
This proof has been mechanized in Coq.
\end{proof}

\subsection*{Theorem \ref{thm:kripkesound}.}
If $\Gamma \proves \phi$ and $K,w,v \models \Gamma$, then $K,w,v \models \phi$.
\begin{proof}
By induction on the derivation of $\Gamma \proves \phi$.
This proof has been mechanized in Coq.
\end{proof}

\subsection*{Proposition \ref{thm:beliefmonotonicity}.}
If $B,w,v \models \phi$ and $w \leq w'$ then $B,w',v \models \phi$.
\begin{proof}
By structural induction on $\phi$.  
This proof has been mechanized in Coq.
\end{proof}

\end{document}